\documentclass[runningheads]{llncs}
\usepackage{graphicx}
\usepackage{amsmath}

\newcommand*{\Perm}[2]{{}^{#1}\!P_{#2}}%

\begin{document}
\title{Self-supervised Denoising via Diffeomorphic Template Estimation: Application to Optical Coherence Tomography}

\titlerunning{Self-supervised OCT Denoising via Diffeomorphic Template Estimation}

\author{Guillaume Gisbert\inst{1} \and Neel Dey\inst{1} \and Hiroshi Ishikawa\inst{2} \and Joel Schuman\inst{2} \and James Fishbaugh\inst{1} \and Guido Gerig\inst{1}}
\index{Gisbert, Guillaume}
\index{Dey, Neel}
\index{Ishikawa, Hiroshi}
\index{Schuman, Joel}
\index{Fishbaugh, James}
\index{Gerig, Guido}
\institute{Computer Science and Engineering, New York University, Brooklyn, NY, USA. \and Ophthalmology, New York University, New York, NY, USA.
\\
\email{neel.dey@nyu.edu}}

\authorrunning{G. Gisbert, et al.}

\maketitle

\begin{abstract}
Optical Coherence Tomography (OCT) is pervasive in both the research and clinical practice of Ophthalmology. However, OCT images are strongly corrupted by noise, limiting their interpretation. Current OCT denoisers leverage assumptions on noise distributions or generate targets for training deep supervised denoisers via averaging of repeat acquisitions. However, recent self-supervised advances allow the training of deep denoising networks using only repeat acquisitions \emph{without} clean targets as ground truth, reducing the burden of supervised learning. Despite the clear advantages of self-supervised methods, their use is precluded as OCT shows strong structural deformations even between sequential scans of the same subject due to involuntary eye motion. Further, direct nonlinear alignment of repeats induces correlation of the noise between images. In this paper, we propose a joint diffeomorphic template estimation and denoising framework which enables the use of self-supervised denoising for motion deformed repeat acquisitions, without empirically registering their noise realizations. Strong qualitative and quantitative improvements are achieved in denoising OCT images, with generic utility in any imaging modality amenable to multiple exposures.

\end{abstract}
\section{Introduction}

\begin{figure}[t]
\centering
\includegraphics[width=0.8\textwidth]{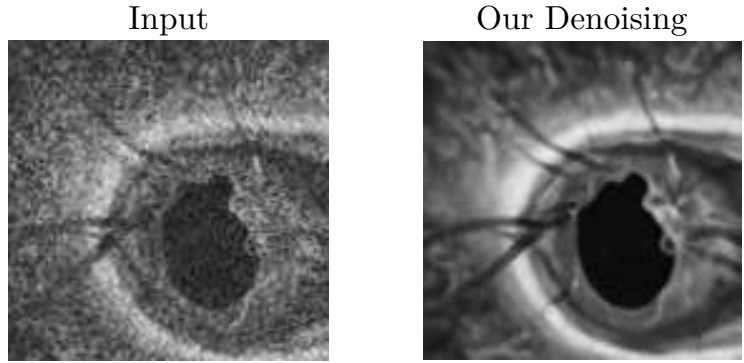}
\caption{Through a joint diffeomorphic template estimation and denoising framework, our methods enable the improved assessment of relevant image features in OCT.} \label{fig:highlight}
\end{figure}

Optical coherence tomography (OCT) is a frontline tool in the non-invasive investigation of ocular microstructure and widely informs critical decisions in Ophthalmology. However, due to the physics of its acquisition, speckle noise permeates OCT volumes and strongly limits the delineation of image structure and its signal-to-noise ratio, a problem which is especially pronounced in images acquired on clinical scanners. To this end, image restoration via denoising may be crucial in the evaluation of OCT imaging, both in a clinical setting and in enabling improved downstream analysis (e.g., segmentation of structures or detection of pathology) in research.

Denoising is a foundational task in image analysis and is an active field of research. Recent years have seen the widespread adoption of supervised deep networks which train mappings between noisy images to clean ones. In OCT or Fluorescence Microscopy, these clean targets for supervised training are often obtained by averaging multiple images \cite{zhang2019poisson,halupka2018retinal,devalla2019deep,qiu2020noise}. Yet, techniques based on dictionary learning \cite{kafieh2014three} or non-local patch statistics in image or transform domains (e.g., BM3D \cite{dabov2007image}) remain competitive in unsupervised image denoising.

Unsupervised deep denoisers have seen rapid progress of late, and have a much lower burden of data preparation as they do not require clean targets. Cycle-consistent image translation methods \cite{zhu2017unpaired} have been used successfully for OCT image enhancement when a dataset acquired on a high-quality scanner is available \cite{romo2020reducing}. However, mappings produced by cycle-consistent methods are brittle and are subject to high-frequency adversarial noise \cite{chu2017cyclegan,bashkirova2019adversarial}. In parallel, several works \cite{krull2019noise2void,batson2019noise2self,laine2019high} show that noisy images can be denoised via masking of the reconstruction loss function or network receptive field. However, this family of methods is inapplicable to OCT due to their strong requirement of independent identically distributed (i.i.d.) pixel noise \cite{hendriksen2020noise2inverse}, leading to checkerboard artefacts.

Most promising for our application, Noise2Noise \cite{Lehtinen2018} makes no pixelwise i.i.d. assumption, and furthermore no assumption on noise following a specific distribution (e.g. Gaussian), and that a denoiser can be trained by replacing the clean target with repeat acquisitions with independent noise between the scans. However, due to eye motion in OCT, repeat scans are structurally misaligned and Noise2Noise is precluded. Even after affine alignment, a linear average of the registered images may be blurry due to nonlinear deformation of ocular microstructure. Correction for this nonlinear deformation is non-trivial as any nonlinear registration algorithm may register noise in addition to structure, breaking the assumptions of Noise2Noise. To this end, a registration and Noise2Noise method was presented in \cite{buchholz2019cryo}, but was designed for only two repeats, and did not address the registration of structure versus noise.

In this paper, we propose a joint diffeomorphic template estimation and denoising framework. Given a subject with $n$ repeats, we first construct a diffeomorphic template \cite{joshi2004unbiased,avants2010optimal,avants2004geodesic} for each subject that minimizes geometric deformation to each of the $n$ repeats while registering each acquisition to this template via careful unsupervised pre-filtering and multi-resolution registration. Once these deformation fields are obtained on the pre-filtered images, we apply these warps to the original images, which empirically ensures that only structure is registered and not noise. We then train a Noise2Noise network on paired slices to denoise the individual OCT images. The presented work is a substantial extension of our preliminary conference abstract \cite{gisbert2020denoising}, including methodological details and thorough experiments. Our approach leads to quantitative improvement of reference-free scores of image quality against several unsupervised denoising methods.

\section{Methods}
\begin{figure}[t]
\includegraphics[width=\textwidth]{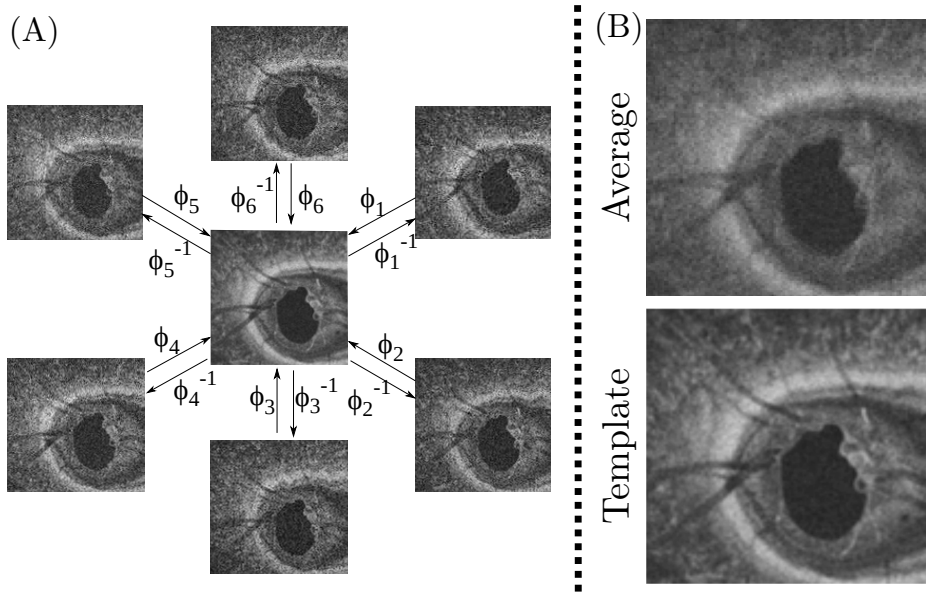}
\caption{(A) A high-level overview of template estimation. (B) Comparison between a linear average after affine alignment (top) and the estimated template (bottom). As templates are deformable averages, they create intrinsically sharper representations.} \label{fig:atlas_building}
\end{figure}

\subsection{Problem formulation}
\begin{figure}[t]
\includegraphics[width=\textwidth]{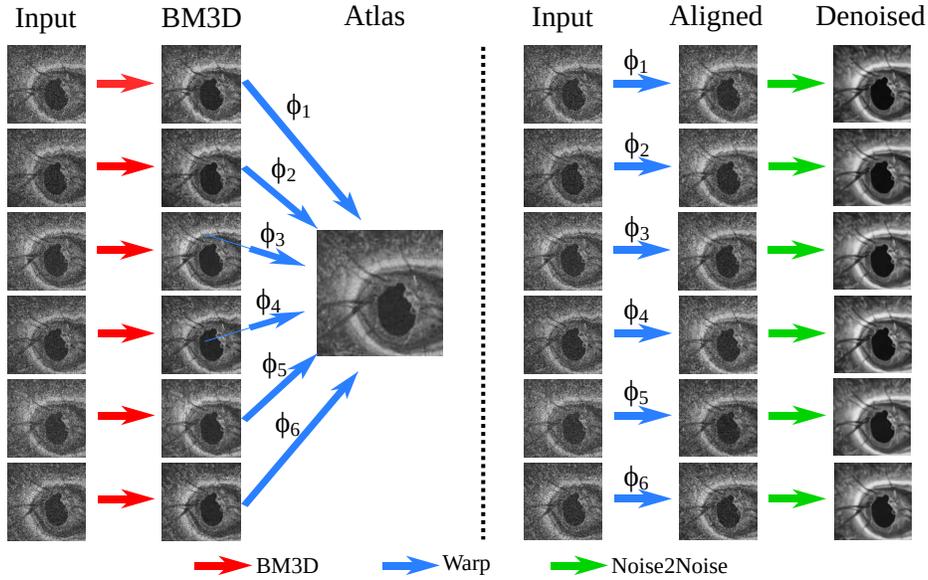}
\caption{Overview of the framework. The raw 3D OCT scans are denoised using BM3D and are then used to build a template with deformations $\Phi_i$. Once obtained, each $\Phi_i$ is applied on its corresponding raw scan, thus enabling the use of Noise2Noise denoising. Deformation fields and templates are estimated in 3D to fully accommodate eye motion.} \label{fig:pipeline_overview}
\end{figure}

A high-level overview of our pipeline is given in Figure \ref{fig:pipeline_overview}. Given a dataset of $m$ subjects with $n$ repeats each with nonlinear deformation, the goal is to learn a denoising function $f$ that maps noisy images to clean images.

To enable the learning of $f$ via a Noise2Noise-like method, we co-register the $n$ repeat 3D OCT scans for each of the $m$ subjects to $m$ subject-specific templates as detailed in Section \ref{sec:registration}. Once registered, $f$ can be efficiently learned via methods detailed in Section \ref{sec:denoising} on 2D slices rather than 3D volumes given the strong anisotropy of OCT imaging.

\subsection{Registration} \label{sec:registration}
Given $n$ repeats to be registered, a reference volume could be arbitrarily chosen to register the remaining $n - 1$ volumes to. However, this approach leads to biased registration estimation  as the user-selected target may require some images to deform significantly more than others depending on the choice of reference. Instead, one can estimate an unbiased template/atlas \cite{joshi2004unbiased,avants2010optimal} which minimizes geometric deformation to each of the $n$ repeats as shown in Figure \ref{fig:atlas_building}(a), doing this by alternating between template estimation and nonlinear registration of the individual images. Of note, template estimation is ubiquitous in neuroimaging, but is starting to find applications in Ophthalmology in both OCT \cite{ravier2018analysis} and retinal fluorescence microscopy \cite{dey2019robust}.

To avoid the registration of noise in addition to structure, each OCT volume is first denoised via BM3D \cite{dabov2007image} applied slicewise. While BM3D does not preserve all structure, it retains sufficient clarity of fine structure for detail preservation in template estimation. We observe that the volumetric equivalent of BM3D (BM4D \cite{maggioni2012nonlocal}) leads to strong block-like artefacts due to the high anisotropy of OCT as shown in Figure \ref{fig:results}. For non-linear registration, we take a multi-resolution strategy, at each stage smoothing with a Gaussian kernel to avoid aliasing, which further encourages the registration of local and global structure rather than local noise. All deformation fields are diffeomorphic, ensuring that no topological changes (tearing, holes, etc.) are made to the structures during registration. This process leads to sharper estimates of image averages as shown in Figure \ref{fig:atlas_building}(b).

Once the deformation fields $\Phi_i$ mapping images to the template are obtained, we discard the prefiltered images and apply the deformations to the original images such that primarily structure, rather than noise, is aligned. We note that this procedure does not theoretically guarantee that no noise is correlated in the registration. However, we find it to be empirically successful towards the desired outcome in our application.

\subsection{Denoising} \label{sec:denoising}
Noise2Noise \cite{Lehtinen2018} is built on the assumption that if a denoising network is trained on repeat images which vary only in noise and not in structure, the clean target can be replaced by a noisy repeat if the noise is zero mean. However, if Noise2Noise is directly applied on affinely aligned images, the output is blurred as shown in Figure \ref{fig:results} as the network cannot distinguish between noise and structure. Once $m$ subject-specific templates for $m$ subjects are built, random pairs of repeat slices from each of the $n$ repeats are extracted and used as a training set. This leads to $(m \times \Perm{n}{2} \times z)$ training points for Noise2Noise where $z$ is the number of 2D slices. As the variance of the denoising estimate decays with the square of the number of training points \cite{Lehtinen2018}, we find that a small OCT dataset with multiple repeats trained slicewise rapidly becomes sufficient for training.

\section{Experiments}
\subsection{Dataset}
Our training dataset consists of 24 subjects who underwent 6 repeat OCT acquisitions, each. The images were of resolution $200 \times 200 \times 600$ were captured on the Cirrus HD5000. Once registered with the process detailed in \ref{sec:registration}, we build every pair of images to train a Noise2Noise network. We crop each slice to a $128 \times 128$ central field of view. Given 6 repeats, we have $\Perm{6}{2}$ pairs for each $z$-slice for each subject. In total, this generates 432,000 training examples. No data augmentation was used in our experiments. As this work pertains to unsupervised denoising without available ground truth, we do not consider a held-out test set.

\subsection{Implementation Details}
Initial pre-registration denoising was performed on 2D slices with a \verb|MATLAB| BM3D implementation\footnote{\url{http://www.cs.tut.fi/~foi/GCF-BM3D/}} with default parameters and $\sigma=0.07$. To estimate subject-specific diffeomorphic templates we use \verb|ANTs|\footnote{\url{http://stnava.github.io/ANTs/}}, with the local normalized cross-correlation metric in a multi-resolution fashion for $50 \times 25 \times 10 \times 10$ iterations at one-sixth, one-fourth, half, and full resolution, respectively.

For training the Noise2Noise-network, we use the \verb|TensorFlow| implementation provided by the authors\footnote{\url{https://github.com/NVlabs/noise2noise}}. Briefly, it employs a U-Net architecture \cite{ronneberger2015u} with 5 downsampling and upsampling layers each. As in the original work, we do not use any regularization or normalization during training. The network was trained for two hundred epochs with a batch size of 4 with the Adam optimizer \cite{kingma2014adam} with an $L_1$ loss, a learning rate of 0.0002 and $\beta_1=0.9$ and $\beta_2=0.999$. Training was performed for two days on a single NVIDIA V100 GPU. New, unseen images are denoised in seconds.

\begin{table}[t]
\begin{center}
\begin{tabular}{|c|c|c|}
\hline
\textbf{Method} & \textbf{Mean Q-metric} \cite{zhu2010automatic} ($\uparrow$) & \textbf{Mean AD} \cite{kong2013new} ($\uparrow$) \\
\hline
Linear Average (affine alignment) & 2.8527 & 0.4677 \\
\hline
Non-Local Means \cite{buades2005non} & 12.2789 & 0.6585 \\
\hline
BM3D (slicewise) \cite{dabov2007image} & 15.7812 & 0.6499 \\
\hline
BM4D \cite{maggioni2012nonlocal} & 12.2843 & 0.6858 \\
\hline
Noise2Noise (affine alignment) \cite{Lehtinen2018} & 16.7860 & 0.4646 \\
\hline
Ours & \textbf{29.5722} & \textbf{0.7183} \\
\hline
\end{tabular}
\label{OVtable}
\end{center}
\caption{Quantitative denoising performance benchmark based on \cite{zhu2010automatic,kong2013new} of all unsupervised denoising methods compared. Higher is better for each score. The linear average and Noise2Noise methods use images after affine alignment but without diffeomorphic atlas registration.}
\end{table}

\subsection{Evaluation methods} \label{sec:eval}
Denoising methods and image quality improvements are popularly benchmarked by using peak signal-to-noise ratio (PSNR) or structural similarity (SSIM) \cite{wang2004image}. However, these scores require noise-free reference images which do not exist in our applications. Further, in preliminary experiments, we found scores that rely on statistics of natural images \cite{mittal2012no} do not correlate well with human perception of denoising quality in OCT images and hence do not use them. Instead, we present two different scores that do not require reference images or training sets of high-quality images.

The first is the Q-metric \cite{zhu2010automatic}, which selects  anisotropic patches in the noisy input to attribute a score based on the singular value decomposition properties of the patch and then averages the patch scores. The second is another no-reference metric developed by \cite{kong2013new} (hence referred to as AD), which measures the structural similarity between the noisy input and the denoised estimation, with the underlying assumption that the noise should be independent in the original image. Higher is better for each score\footnote{For \cite{kong2013new}, Algorithm 1 in the paper suggests that lower is better. However, their code negates the final correlation value, thus making higher better. We do the same to maintain consistency with their convention.}. Of note, both of these scores are image-dependent, i.e., score improvement on one image cannot be compared to another image. However, we average the scores on the same dataset for all denoising methods benchmarked, so the values are now comparable.

\begin{figure}[!t]
\centering
\includegraphics[width=0.8\textwidth]{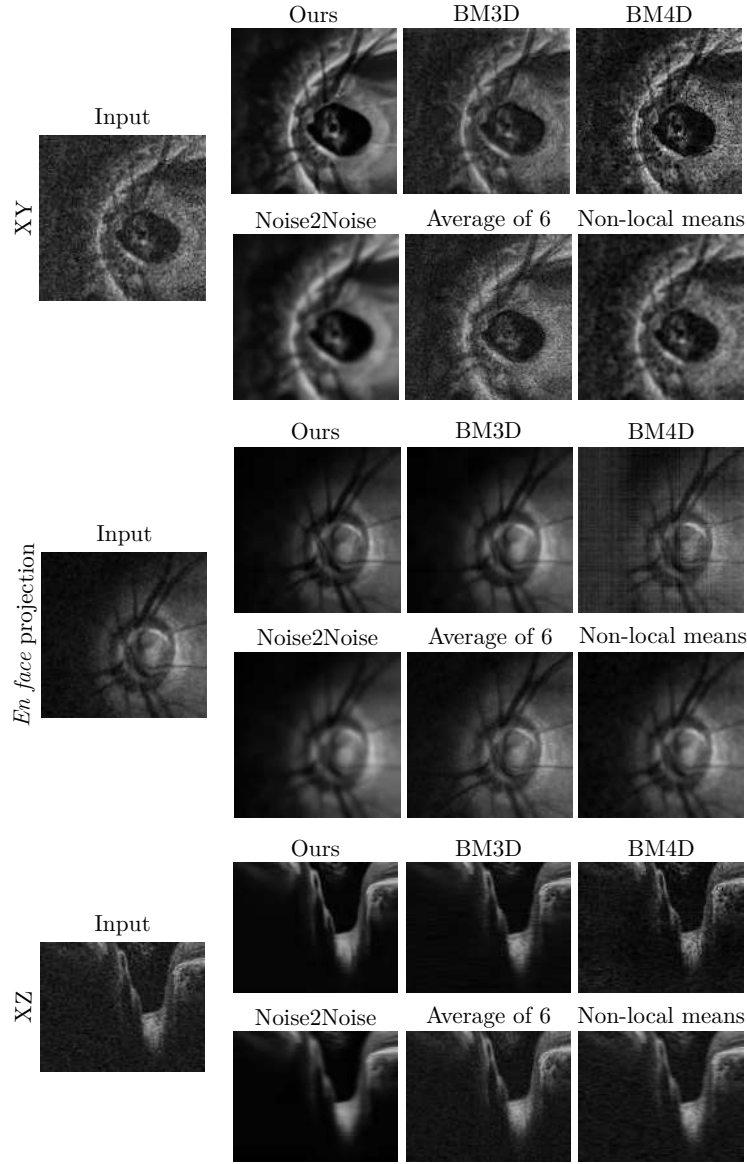}
\caption{A qualitative comparison of denoising quality of all benchmarked methods. Our network was trained and tested on XY-slices only, with the other views assembled and included for completeness. Readers are encouraged to zoom-in for details. } \label{fig:results}
\end{figure}

\subsection{Results} \label{sec:results}

We benchmark a variety of unsupervised denoising methods, including a simple linear average of all repeats after affine alignment, Non-Local Means \cite{buades2005non}, BM3D \cite{dabov2007image}, BM4D \cite{maggioni2012nonlocal}, Noise2Noise (with affine alignment only), and our method. We do not compare with self-supervised methods that require only single noisy images without pairs \cite{krull2019noise2void,batson2019noise2self,laine2019high} as these methods require pixelwise i.i.d. noise not satisfied by OCT images, thus leading to checkerboard artefacts \cite{hendriksen2020noise2inverse} and creating an unfair comparison.

Qualitative results are shown in Figure \ref{fig:results} for the XY slices that our model was trained on, further including assembled XZ and \textit{en face} mean projections commonly used in ophthalmology for completeness. We observe that while a linear average does denoise, subtle details are lost due to nonlinear deformation. Non-local 2D patch-based methods blur details that may be relevant structurally. BM4D introduces significant block-like artefacts visible in the \textit{en face} view due to image anisotropy. A direct application of Noise2Noise on affinely aligned images significantly blurs detail. Finally, our proposed method shows a net improvement in image quality, reducing noise drastically while preserving sharp edges.

Quantitatively, we report the dataset-wide scores detailed in Table \ref{OVtable}. We find that for both scores, the proposed framework outperforms the baselines. Interestingly, we note that the blurry reconstruction produced by a direct application of Noise2Noise (with affine alignment) is marginally preferred by the Q-metric over the slicewise application of BM3D (which is perceptually better denoised). This may suggest that AD is a more reliable score of image denoising quality for this application, as pointed out by \cite{kong2013new}.

\section{Discussion}
In this paper, we present a self-supervised framework to denoise repeat acquisitions of images subject to strong noise and deformation, common in OCT imaging. Strong qualitative and quantitative improvements are observed w.r.t. unsupervised baselines, denoising while maintaining fine detail and allowing for clearer morphological interpretation. Furthermore, as OCT practice often averages multiple scans to remove noise, repeated scans are typically available. Lastly, the method is generic and could be applied to other imaging modalities which are subject to deformation and noise in repeat acquisitions, e.g., live fluorescence microscopy.

While repeat observations are required for training, the model can be directly applied to unseen individual scans from new subjects thereafter. This enables the denoising of large amounts of retrospective data, as long as there is no significant domain gap. 

In future work, we will investigate the ability of deep self-supervised denoisers that require no repeat acquisitions in handling the correlated noise typical to OCT. Very recent work \cite{broaddus2020removing} develops such a method for handling structured noise in fluorescence microscopy, but requires a heuristic estimation of the structure of noise. Finally, our framework is not amenable to end-to-end training of registration and denoising, and such a joint method may have improved results and widespread utility in biomedical imaging.
\\
\\
\noindent\textbf{Acknowledgments} \quad This work was supported by NIH grants 1R01EY027948-01 and 2R01EY013178-15. HPC resources used for this research provided by grant NSF MRI-1229185.

%
%
\bibliographystyle{splncs04}
\bibliography{ref}

\begin{thebibliography}{10}
\providecommand{\url}[1]{\texttt{#1}}
\providecommand{\urlprefix}{URL }
\providecommand{\doi}[1]{https://doi.org/#1}

\bibitem{avants2004geodesic}
Avants, B., Gee, J.C.: Geodesic estimation for large deformation anatomical
  shape averaging and interpolation. Neuroimage  \textbf{23},  S139--S150
  (2004)

\bibitem{avants2010optimal}
Avants, B.B., Yushkevich, P., Pluta, J., Minkoff, D., Korczykowski, M., Detre,
  J., Gee, J.C.: The optimal template effect in hippocampus studies of diseased
  populations. Neuroimage  \textbf{49}(3),  2457--2466 (2010)

\bibitem{bashkirova2019adversarial}
Bashkirova, D., Usman, B., Saenko, K.: Adversarial self-defense for
  cycle-consistent gans. In: Advances in Neural Information Processing Systems.
  pp. 637--647 (2019)

\bibitem{batson2019noise2self}
Batson, J., Royer, L.: Noise2self: Blind denoising by self-supervision. In:
  International Conference on Machine Learning. pp. 524--533 (2019)

\bibitem{broaddus2020removing}
Broaddus, C., Krull, A., Weigert, M., Schmidt, U., Myers, G.: Removing
  structured noise with self-supervised blind-spot networks. In: 2020 IEEE 17th
  International Symposium on Biomedical Imaging (ISBI). pp. 159--163. IEEE
  (2020)

\bibitem{buades2005non}
Buades, A., Coll, B., Morel, J.M.: A non-local algorithm for image denoising.
  In: 2005 IEEE Computer Society Conference on Computer Vision and Pattern
  Recognition (CVPR'05). vol.~2, pp. 60--65. IEEE (2005)

\bibitem{buchholz2019cryo}
Buchholz, T.O., Jordan, M., Pigino, G., Jug, F.: Cryo-care: content-aware image
  restoration for cryo-transmission electron microscopy data. In: 2019 IEEE
  16th International Symposium on Biomedical Imaging (ISBI 2019). pp. 502--506.
  IEEE (2019)

\bibitem{chu2017cyclegan}
Chu, C., Zhmoginov, A., Sandler, M.: Cyclegan, a master of steganography. arXiv
  preprint arXiv:1712.02950  (2017)

\bibitem{dabov2007image}
Dabov, K., Foi, A., Katkovnik, V., Egiazarian, K.: Image denoising by sparse
  3-d transform-domain collaborative filtering. IEEE Transactions on image
  processing  \textbf{16}(8),  2080--2095 (2007)

\bibitem{devalla2019deep}
Devalla, S.K., Subramanian, G., Pham, T.H., Wang, X., Perera, S., Tun, T.A.,
  Aung, T., Schmetterer, L., Thi{\'e}ry, A.H., Girard, M.J.: A deep learning
  approach to denoise optical coherence tomography images of the optic nerve
  head. Scientific reports  \textbf{9}(1),  1--13 (2019)

\bibitem{dey2019robust}
Dey, N., Messinger, J., Smith, R.T., Curcio, C.A., Gerig, G.: Robust
  non-negative tensor factorization, diffeomorphic motion correction, and
  functional statistics to understand fixation in fluorescence microscopy. In:
  International Conference on Medical Image Computing and Computer-Assisted
  Intervention. pp. 658--666. Springer (2019)

\bibitem{gisbert2020denoising}
Gisbert, G., Dey, N., Ishikawa, H., Schuman, J., Fishbaugh, J., Gerig, G.:
  Improved denoising of optical coherence tomography via repeated acquisitions
  and unsupervised deep learning. Investigative Ophthalmology \& Visual Science
   \textbf{61}(9),  PB0035 (2020)

\bibitem{halupka2018retinal}
Halupka, K.J., Antony, B.J., Lee, M.H., Lucy, K.A., Rai, R.S., Ishikawa, H.,
  Wollstein, G., Schuman, J.S., Garnavi, R.: Retinal optical coherence
  tomography image enhancement via deep learning. Biomedical optics express
  \textbf{9}(12),  6205--6221 (2018)

\bibitem{hendriksen2020noise2inverse}
Hendriksen, A.A., Pelt, D.M., Batenburg, K.J.: Noise2inverse: Self-supervised
  deep convolutional denoising for linear inverse problems in imaging. arXiv
  preprint arXiv:2001.11801  (2020)

\bibitem{joshi2004unbiased}
Joshi, S., Davis, B., Jomier, M., Gerig, G.: Unbiased diffeomorphic atlas
  construction for computational anatomy. NeuroImage  \textbf{23},  S151--S160
  (2004)

\bibitem{kafieh2014three}
Kafieh, R., Rabbani, H., Selesnick, I.: Three dimensional data-driven multi
  scale atomic representation of optical coherence tomography. IEEE
  transactions on medical imaging  \textbf{34}(5),  1042--1062 (2014)

\bibitem{kingma2014adam}
Kingma, D.P., Ba, J.: Adam: A method for stochastic optimization. arXiv
  preprint arXiv:1412.6980  (2014)

\bibitem{kong2013new}
Kong, X., Li, K., Yang, Q., Wenyin, L., Yang, M.H.: A new image quality metric
  for image auto-denoising. In: Proceedings of the IEEE International
  Conference on Computer Vision. pp. 2888--2895 (2013)

\bibitem{krull2019noise2void}
Krull, A., Buchholz, T.O., Jug, F.: Noise2void-learning denoising from single
  noisy images. In: Proceedings of the IEEE Conference on Computer Vision and
  Pattern Recognition. pp. 2129--2137 (2019)

\bibitem{laine2019high}
Laine, S., Karras, T., Lehtinen, J., Aila, T.: High-quality self-supervised
  deep image denoising. In: Advances in Neural Information Processing Systems.
  pp. 6970--6980 (2019)

\bibitem{Lehtinen2018}
Lehtinen, J., Munkberg, J., Hasselgren, J., Laine, S., Karras, T., Aittala, M.,
  Aila, T.: {Noise2Noise}: Learning image restoration without clean data. In:
  ICML. pp. 2971--2980 (2018)

\bibitem{maggioni2012nonlocal}
Maggioni, M., Katkovnik, V., Egiazarian, K., Foi, A.: Nonlocal transform-domain
  filter for volumetric data denoising and reconstruction. IEEE transactions on
  image processing  \textbf{22}(1),  119--133 (2012)

\bibitem{mittal2012no}
Mittal, A., Moorthy, A.K., Bovik, A.C.: No-reference image quality assessment
  in the spatial domain. IEEE Transactions on image processing
  \textbf{21}(12),  4695--4708 (2012)

\bibitem{qiu2020noise}
Qiu, B., Huang, Z., Liu, X., Meng, X., You, Y., Liu, G., Yang, K., Maier, A.,
  Ren, Q., Lu, Y.: Noise reduction in optical coherence tomography images using
  a deep neural network with perceptually-sensitive loss function. Biomedical
  Optics Express  \textbf{11}(2),  817--830 (2020)

\bibitem{ravier2018analysis}
Ravier, M., Hong, S., Girot, C., Ishikawa, H., Tauber, J., Wollstein, G.,
  Schuman, J., Fishbaugh, J., Gerig, G.: Analysis of morphological changes of
  lamina cribrosa under acute intraocular pressure change. In: International
  Conference on Medical Image Computing and Computer-Assisted Intervention. pp.
  364--371. Springer (2018)

\bibitem{romo2020reducing}
Romo-Bucheli, D., Seeb{\"o}ck, P., Orlando, J.I., Gerendas, B.S., Waldstein,
  S.M., Schmidt-Erfurth, U., Bogunovi{\'c}, H.: Reducing image variability
  across oct devices with unsupervised unpaired learning for improved
  segmentation of retina. Biomedical Optics Express  \textbf{11}(1),  346--363
  (2020)

\bibitem{ronneberger2015u}
Ronneberger, O., Fischer, P., Brox, T.: U-net: Convolutional networks for
  biomedical image segmentation. In: International Conference on Medical image
  computing and computer-assisted intervention. pp. 234--241. Springer (2015)

\bibitem{wang2004image}
Wang, Z., Bovik, A.C., Sheikh, H.R., Simoncelli, E.P.: Image quality
  assessment: from error visibility to structural similarity. IEEE transactions
  on image processing  \textbf{13}(4),  600--612 (2004)

\bibitem{zhang2019poisson}
Zhang, Y., Zhu, Y., Nichols, E., Wang, Q., Zhang, S., Smith, C., Howard, S.: A
  poisson-gaussian denoising dataset with real fluorescence microscopy images.
  In: Proceedings of the IEEE Conference on Computer Vision and Pattern
  Recognition. pp. 11710--11718 (2019)

\bibitem{zhu2017unpaired}
Zhu, J.Y., Park, T., Isola, P., Efros, A.A.: Unpaired image-to-image
  translation using cycle-consistent adversarial networks. In: Proceedings of
  the IEEE international conference on computer vision. pp. 2223--2232 (2017)

\bibitem{zhu2010automatic}
Zhu, X., Milanfar, P.: Automatic parameter selection for denoising algorithms
  using a no-reference measure of image content. IEEE transactions on image
  processing  \textbf{19}(12),  3116--3132 (2010)

\end{thebibliography}
\end{document}